\documentclass[aps,prb,twocolumn,superscriptaddress,english,floatfix,longbibliography]{revtex4-2}
\usepackage{graphicx}
\usepackage{float}
\usepackage{physics}
\usepackage{cancel}
\usepackage{overpic}
\usepackage{enumerate}
\usepackage{dsfont}

\usepackage[colorlinks,citecolor=blue,linkcolor=blue,urlcolor=blue]{hyperref}
\usepackage{textcomp}
\usepackage{amsmath}
\usepackage{amssymb}
\usepackage{soul}
\usepackage[normalem]{ulem}
\usepackage{lipsum}
\usepackage{comment}
\usepackage{mathrsfs}

\usepackage[english]{babel}
\usepackage{bm}
\usepackage{orcidlink}

\graphicspath{{./}{figsArticle/}}

\renewcommand{\(}{\left(}
\renewcommand{\)}{\right)}
\renewcommand{\[}{\left[}
\renewcommand{\]}{\right]}

\newcommand{\calpha}{\alpha^\ast}

\renewcommand{\hl}[1]{{\color{black} #1}}

\usepackage{tcolorbox} 
\newtcolorbox{redbox}[1]{colback=red!5!white,colframe=red!75!black,fonttitle=\bfseries,title=#1}
\newtcolorbox{blackbox}[1]{colback=black!5!white,colframe=black!75!black,fonttitle=\bfseries,title=#1}

\newtcolorbox{bluebox}[1]{colback=blue!5!white,colframe=blue!75!black,fonttitle=\bfseries,title=#1}

\let\originalleft\left
\let\originalright\right
\renewcommand{\left}{\mathopen{}\mathclose\bgroup\originalleft}
\renewcommand{\right}{\aftergroup\egroup\originalright}

\newcommand{\vect}[1]{\boldsymbol{#1}}
\renewcommand{\vec}[1]{\vect{#1}}

\makeatletter
\newcommand*\bigcdot{{\color{gray}\mathpalette\bigcdot@{1.}}}
\newcommand*\bigcdot@[2]{\mathbin{\vcenter{\hbox{\scalebox{#2}{$\m@th#1\bullet$}}}}}
\makeatother

\def\rme{{\rm {e}}}
\def\rmi{{\rm {i}}}
\renewcommand{\d}{{\rm {d}}}

\newcommand{\idhat}{\hat{\mathds{1}}}

\def\tr{{\rm{Tr}}}
\newcommand{\abss}[1]{\left|#1\right|^2}

\newcommand{\alphatil}{{\Tilde{\alpha}}}

\newcommand{\delhat}{\hat{\delta}}
\newcommand{\ddelhat}{\hat{\delta}^\dagger}

\newcommand{\rhohat}{\hat{\rho}}

\newcommand{\schr}{Schr\"{o}dinger}

\newcommand{\hc}{\mathrm{H.c.}}

\newcommand{\aaa}{\hat{a}}
\newcommand{\daaa}{\hat{a}^\dagger}
\newcommand{\daaas}{\hat{a}^{\dagger 2}}
\newcommand{\aaas}{\aaa^{2}}

\newcommand{\acal}{\mathcal{A}}

\newcommand{\bbb}{\hat{b}}
\newcommand{\dbbb}{\hat{b}^\dagger}

\newcommand{\ccal}{\mathcal{C}}

\newcommand{\ecal}{\mathcal{E}}

\newcommand{\fcal}{\mathcal{F}}
\newcommand{\gcal}{\mathcal{G}}

\newcommand{\ecalf}{\ecal^{F}}
\newcommand{\Ftil}{\tilde{F}}

\newcommand{\ecald}{\ecal^{D}}
\newcommand{\fcald}{\fcal^{D}}
\newcommand{\czd}{\ccal_{\ztil}^{D}}

\newcommand{\ecaldc}{\ecal^{D*}}
\newcommand{\fcaldc}{\fcal^{D*}}
\newcommand{\czdc}{\ccal_{\ztil}^{D*}}

\newcommand{\hcal}{\mathcal{H}}

\newcommand{\Hhat}{\hat{H}}

\newcommand{\Jcal}{\mathcal{J}}

\newcommand{\Khat}{\hat{K}}

\newcommand{\lcal}{\mathcal{L}}
\newcommand{\Lhat}{\hat{L}}

\newcommand{\Lcal}{\mathcal{L}}

\newcommand{\mvec}{{\vec{m}}}

\newcommand{\nh}{{\mathrm{nh}}}
\newcommand{\ohat}{\hat{o}}

\newcommand{\Ocal}{\mathcal{O}}

\newcommand{\rf}{{\mathrm{RF}}}

\newcommand{\Shat}{\hat{S}}

\newcommand{\Uhat}{\uhat}
\newcommand{\uhat}{\hat{U}}

\newcommand{\Util}{\tilde{U}}

\newcommand{\Ucal}{\mathcal{U}}

\newcommand{\xtil}{{\Tilde{x}}}

\newcommand{\ytil}{{\Tilde{y}}}

\newcommand{\ztil}{{\Tilde{z}}}

\newcommand{\pmx}[1]{\begin{pmatrix}
		#1
\end{pmatrix}}

\newcommand{\bea}{\begin{equation}\begin{aligned}}
		\newcommand{\eea}{\end{aligned}\end{equation}}
\newcommand{\be}{\begin{equation}}
	\newcommand{\ee}{\end{equation}}

\renewcommand{\selectlanguage}[1]{}
\begin{document}

	\title{Emergent deterministic entanglement dynamics in monitored infinite-range bosonic systems}
	
	\author{Zejian Li~\orcidlink{0000-0002-5652-7034}}
	\affiliation{The Abdus Salam International Center for Theoretical Physics, Strada Costiera 11, 34151 Trieste, Italy}
	\author{Anna Delmonte~\orcidlink{0009-0008-9371-6855}}
	\affiliation{SISSA, Via Bonomea 265, I-34136 Trieste, Italy}

	\author{Rosario Fazio~\orcidlink{0000-0002-7793-179X}}
	\affiliation{The Abdus Salam International Center for Theoretical Physics, Strada Costiera 11, 34151 Trieste, Italy}
	\affiliation{Dipartimento di Fisica ``E. Pancini", Universit\`a di Napoli ``Federico II'', Monte S. Angelo, I-80126 Napoli, Italy}
	
	\begin{abstract}
		We study monitored quantum dynamics of infinite-range interacting bosonic systems in the thermodynamic limit. We show that under semiclassical assumptions, the quantum fluctuations along single monitored trajectories adopt a deterministic limit for both quantum-jump and state-diffusion unravelings, and can be exactly solved. In particular, the hierarchical structure of the equations of motion explains the coincidence of entanglement \hl{criticalities} and dissipative phase transitions found in previous finite-size numerical studies. We illustrate the findings on a Bose-Hubbard dimer and a collective spin system.

	\end{abstract}
	\maketitle

	\section{ Introduction}
The dynamics of monitored quantum many-body systems has recently triggered intense interest in both theoretical and experimental activities~\cite{Skinner_prx,Li_Fisher,noel2021measurementinducedquantum,KohIBM,Google1}. When a quantum system is continuously measured, the state of the system evolves along a stochastic \textit{quantum trajectory} conditioned by the measurement outcomes, resulting from the interplay between its unitary dynamics and the interaction with the external environment. \hl{When averaged over all possible measurement outcomes, the mixed state typically admits a deterministic evolution governed by a master equation that describes the dissipative dynamics of the system.} It was realized in recent years that the quantum trajectory ensemble of a monitored many-body system contains rich and exotic phenomenology with no counterpart in equilibrium or dissipative settings. \hl{Pioneer examples are} measurement-induced phase transitions~\cite{Skinner_prx,Li_Fisher} \hl{found in random unitary circuits interspersed with local projective measurements, where the system undergoes a transition from a volume-law entangled phase to an area-law (disentangled) phase upon increasing the measurement rate},  an emergent critical phenomena visible only on the level of quantum trajectories through nonlinear probes of the conditioned state, such as entanglemen.
\hl{ The mixed state evolution, on the other hand, generally admits different dynamics independent of the monitored phases, such as being completely featureless~\cite{Skinner_prx,Li_Fisher} or showing dissipative phase transitions at different critical points~\cite{piroli2022trivialityofquantum,sierant2023controlling,sierant2022dissipative}. A large body of initial investigations on monitored many-body phases has focused on circuit models~\cite{ippoliti2021entanglementphasetransitions,Sierant_2023,sierant2023controlling,Kalsi_Romito_Schomerus_2022,ware2023sharp} and also in Hamiltonian settings~\cite{BucchholdFreeF,BuchholdDirac,BuchholdFreeB,FujiAshida,Cao,Kells,Turkeshi,sierant2022dissipative,paviglianiti2023multipartiteentanglementin,guido2,Tirrito_Santini_Fazio_Collura_2023} for weakly monitored systems, including settings with nonlocal interactions~\cite{SwingleSYk,Block,sharmaMeasurementinducedCriticalityExtended2022}, bringing deep insights to the nature of measurement-induced phases, for example, in their connection to error-correction properties of quantum channels~\cite{Gullans}.}

	Out of the various scenarios studied in the context of monitored phases, infinite-range interacting systems stood out by their entanglement dynamics favoring the experimental detection of entanglement phase transitions. In Refs.~\cite{Passarelli,delmonteMeasurementinducedPhaseTransitions2025}, some of us studied infinite-range spins and bosonic systems, showing that the entanglement features fast-saturating dynamics, thereby mitigating the experimental overhead in detecting the entanglement phases via post-selection~\cite{gullans2020scalableprobesof,Ippoliti}. 
	The findings of these works share a common peculiarity: the average state of the system, undergoing dissipative Lindblad dynamics, also exhibits a phase transition, which always emerges at the same critical point as the entanglement phase transition, as suggested by finite-size numerics. It is worth understanding, from a theoretical perspective, whether this feature is intrinsic to infinite-range interacting systems.

	In this \hl{paper}, we complete the previous results by studying the monitored dynamics of infinite-range bosonic systems directly in the thermodynamic limit. In the particular case of free bosons with state-diffusion type of monitoring of linear field operators, it has been shown by Y. Minoguchi, P. Rabl, and M. Buchhold in Ref.~\cite{BuchholdFreeB} that the entanglement follows the same deterministic evolution along any monitored trajectory. This arises from the fact that the system state remains Gaussian under non-interacting dynamics, which can be fully characterized by its first moments and the covariance matrix, with the latter being deterministic and defining the entanglement. In the more general case of infinite-range interacting bosons, the system can be well approximated by a Gaussian state due to its semiclassical nature~\cite{buchholzMultivariateCentralLimit2014,goderisCentralLimitTheorem1989,verbeureManyBodyBosonSystems2011}, and the theory becomes asymptotically mean-field and non-interacting~\cite{carolloExactnessMeanFieldEquations2021,carolloApplicabilityMeanFieldTheory2024a,mattesLongrangeInteractingSystems2024}, allowing us to study the dissipative dynamics exploiting the Gaussian framework. When monitoring is involved, this formalism typically results in stochastic equations of the Gaussian parameters along quantum trajectories~\cite{verstraelen2018gaussian}, which is the numerical approach we previously adopted in studying infinite-range Bosons~\cite{delmonteMeasurementinducedPhaseTransitions2025} and long-range spins (with bosonized spin-wave fluctuations)~\cite{liMonitoredLongrangeInteracting2025a}. Based on the Gaussian-state approximations, we make in this letter the key observation that the stochasticity in the monitored semiclassical dynamics (for both quantum-jump and state-diffusion processes) is suppressed by system size, leading to deterministic equations for the covariances in the thermodynamic limit, whose dynamics is driven by the standard mean-field solutions for the first moments. In particular, this allows the analytical study of entanglement dynamics along monitored trajectories and explains the coincidence of entanglement transitions and symmetry-breaking dissipative transitions.
	We illustrate these findings on two infinite-range interacting models, a Bose-Hubbard dimer and a collective spin system (which can be mapped to a bosonic model), where we solve the monitored entanglement dynamics exactly in the thermodynamic limit.

	\section{ Monitored infinite-range bosonic system}
	Let us consider an infinite-range bosonic system with annihilation operators $\aaa_i$ representing the collective modes, that satisfy standard bosonic commutation relations $[\aaa_i,\daaa_j]=\delta_{ij}$. The Gaussian-state approximation allows the system state to be fully determined by the first moments $\alpha_i \equiv\langle\aaa_i\rangle$ and the covariances $u_{ij} \equiv\langle\delhat_i\delhat_j\rangle$, $v_{ij} \equiv \langle\ddelhat_i\delhat_j\rangle$, where the fluctuation operators are defined as
	$\delhat_i\equiv\aaa_i-\langle\aaa_i\rangle$. We make the following semiclassical assumption on the scaling of the Gaussian parameters, such that $\alpha_i\sim\Theta(\sqrt{N})$ and $u\sim v\sim\Ocal(1)$,     
	where the parameter $N$ is proportional to the bosonic population and defines the thermodynamic limit via $N\to\infty$. 
	It then follows that the state is approximately coherent up to $\Ocal(1/N)$ corrections, i.e.,
	\bea\label{eq:coherent-approx}
	\langle\aaa_i^{\dagger m}\aaa^n_j\rangle = \alpha^{*m}_i\alpha^n_j[1 + \Ocal(1/N)]\,.
	\eea
	These approximations are typically satisfied in infinite-range interaction systems in the large $N$ limit as a result of quantum central limit theorems~\cite{buchholzMultivariateCentralLimit2014,goderisCentralLimitTheorem1989,verbeureManyBodyBosonSystems2011}.
	
	We consider the quantum trajectory resulting from an unraveled Lindblad dynamics \hl{represented by the following stochastic master equation} (with $\hbar=1$)~\cite{breuer2002theory,Wiseman},
	\bea\label{eq:mf-coarse-general}
	\d\rhohat = \d t
	\lcal\rhohat + \d\hcal \rhohat\,,
	\eea
	where $\lcal$ denotes the Liouvillian superoperator,
	\bea
	\lcal\rhohat\equiv-\rmi[\Hhat,\rhohat] + \sum_i\left(\Lhat_i\rhohat\Lhat_i^\dagger-\dfrac{1}{2}\{\Lhat_i^\dagger\Lhat_i,\rhohat\}\right)\,,
	\eea
	with the Hamiltonian $\Hhat$ and jump operators $\Lhat_i$, 
	and
	\bea
	\d\hcal\rhohat\equiv \sum_i \left[\d Z_i^*(\Lhat_i-\langle\Lhat_i\rangle)\rhohat + \hc\right]
	\eea
	encodes the stochastic terms arising from the monitoring, where $\d Z_i$ is a complex zero-mean noise that we now specify.
	As shown in Appendix~\ref{app:qj-semi}, this represents the general form of monitored dynamics in the semiclassical limit, incorporating both quantum-jump and state-diffusion unravelings, via the definition of the noise variance,
	\bea
	\d Z_i\d Z_j = \Upsilon_i \delta_{ij}\d t\,,\quad
	\d Z_i^*\d Z_j = \delta_{ij}\d t\,,
	\eea
	and the choice of the unraveling is achieved as follows,
	\bea\label{eq:upsilon}
	\Upsilon_i = \begin{cases}
		\dfrac{\langle\Lhat_i\rangle^2}{|\langle\Lhat_i\rangle|^2} &\text{quantum jump,}\\
		1 &\text{homodyne,}\\
		0 & \text{heterodyne.}
	\end{cases}
	\eea
	Intuitively, the case of quantum jump can be understood from the fact that in the semiclassical limit, the jump rate diverges while modification in the state by a single jump is vanishingly small, resulting in the continuous form of a state-dependent general-dyne unraveling when coarse-grained.
	
	\begin{figure*}[ht]
		\centering
		\includegraphics[width=0.9\linewidth]{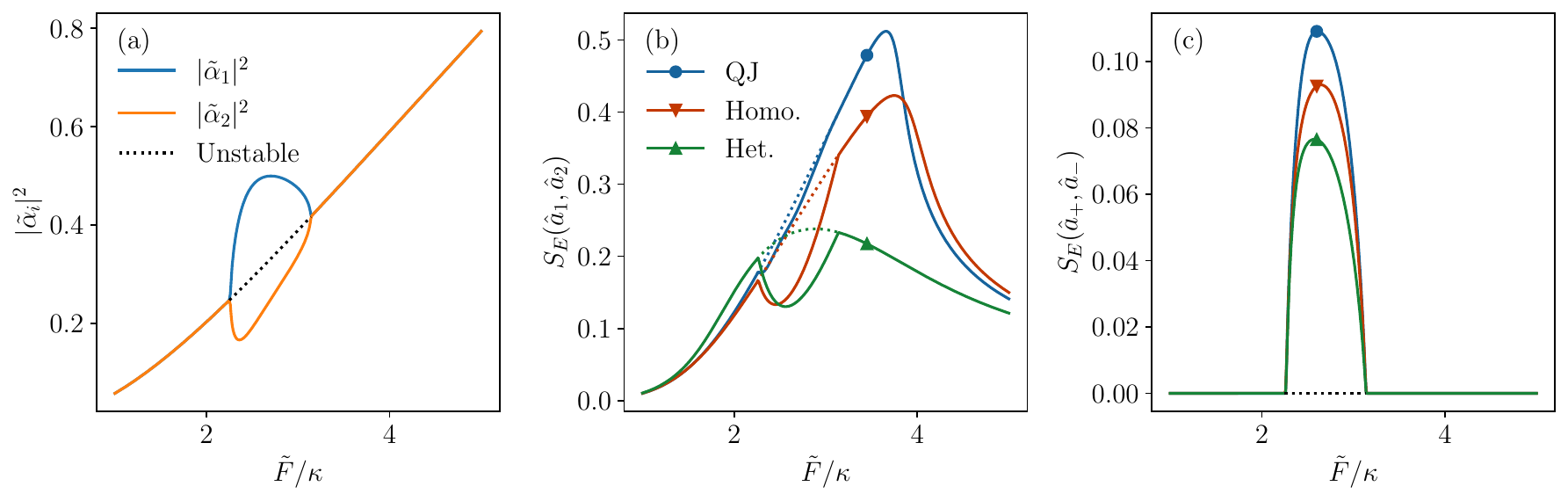}
		\caption{Steady-state features of the Bose-Hubbard dimers with different unravelings in the thermodynamic limit. (a) Normalized population $|\alphatil_i|^2$ in the two cavities (assuming $|\alphatil_1|^2\geq |\alphatil_2|^2$). The solid (dashed) line marks the stable (unstable) solution in all panels. (b) The entanglement entropy between the two spatial modes of the dimer $S_E(\aaa_1,\aaa_2)$ for different unravelings [quantum jump (QJ), homodyne (Homo.) and heterodyne (Het.), see legend]. (c) Same as panel (b) but for the entanglement between the two momentum modes $S_E(\aaa_+,\aaa_-)$ of the dimer. Parameters: $J=2.5\kappa$, $\Delta=-1.5\kappa$, $\Tilde{U}=2\kappa$.}
		\label{fig:ept-bhd-qj-dynes}
	\end{figure*}

	\section{ Deterministic dynamics of quantum fluctuations in the thermodynamic limit}\label{sec:det-fluc}
	We now consider the quantum fluctuation dynamics in the large $N$ limit. To be consistent with the semiclassical assumptions stated above, we require the model to be extensive in $N$, i.e., $\langle\Hhat\rangle\sim \Theta(N)$ and $\langle\Lhat_i\rangle\sim\Theta(\sqrt{N})$, which ensures a well-defined thermodynamic limit. Within the Gaussian state assumption, it then follows that the normalized first moment $\alphatil_i\equiv\alpha_i/\sqrt{N}$ and the covariances $u_{ij}$, $v_{ij}$ follow deterministic dynamics in the $N\to\infty$ limit (see Appendix~\ref{app:det-limit} for a detailed discussion), which can be formally written as follows (\hl{see \hl{Appendix~\ref{app:benchmark-single-kerr}} for an example of a single-mode Bosonic model benchmarked against exact numerics.}),
	
	\bea\label{eq:det-boson-multi}
	\d \alphatil_i &= \left\{\d t\dfrac{1}{\sqrt{N}}\langle\Lcal^\ddagger\aaa_i\rangle\right\}_0\,,\\
	\d u_{ij} &= \left\{\d t\langle\Lcal^\ddagger(\delhat_i\delhat_j)\rangle-\langle\d\hcal^\ddagger\aaa_i\rangle\langle\d\hcal^\ddagger\aaa_j\rangle\right\}_0\,,\\ \d v_{ij} &=\left\{ \d t\langle\Lcal^\ddagger(\ddelhat_i\delhat_j)\rangle-\langle\d\hcal^\ddagger\aaa_i\rangle^*\langle\d\hcal^\ddagger\aaa_j\rangle\right\}_0\,,
	\eea
	where $\{\bullet\}_0$ denotes the terms independent of $N$, and any term beyond quadratic order is to be expressed using the first and second moments using Wick's theorem~\cite{wickEvaluationCollisionMatrix1950}. A hierarchical structure is present in these equations: the equation for $\alphatil_i$ is the standard mean-field equation, which does not involve the covariances; on the other hand, the dynamics of $u_{ij}$ and $v_{ij}$ is driven by the mean-field solution $\alphatil_i(t)$. The terms given by It\^{o} differentiation rules at the end of the equations for $u_{ij}$ and $v_{ij}$ inherit the effect of monitoring and also ensure the purity of the state along the unraveled trajectory.
	
	A direct consequence is that the entanglement, dependent only on the \hl{single-trajectory} covariances~\cite{serafiniSymplecticInvariantsEntropic2003} $u_{ij}$ and $v_{ij}$, also evolves deterministically along every single monitored trajectory in the thermodynamic limit. This formalism thus provides a well-defined mean-field picture for monitored dynamics, allowing for the analytical study of entanglement \hl{criticalities}. Moreover, the hierarchical structure of the equations~\eqref{eq:det-boson-multi} implies that whenever the mean-field theory (equation for $\alphatil_i$) predicts a phase transition in the thermodynamic limit, we also expect non-analytical behavior to occur in the entanglement due to its (typically smooth) dependence on the mean-field solution. 
	
	\section{ Bose-Hubbard dimer}\label{sec:bh-dimer-results}
	Let us now apply the formalism to study a Bose-Hubbard dimer consisting of two coupled Kerr cavities~\cite{casteelsOpticallyBistableDrivendissipative2017,lledoDrivenBoseHubbardDimer2019,pudlikDynamicsEntanglementDissipative2013,casteelsQuantumEntanglementSpatialsymmetrybreaking2017,giraldoSemiclassicalBifurcationsQuantum2022,liDissipationinducedAntiferromagneticlikeFrustration2021},
	\bea\label{eq:bh-ham}
	\Hhat &= \sum_{i=1}^{2}\left[ -\Delta\daaa_i\aaa_i + \dfrac{\Util}{2N}\daaas_i\aaas_i+\Ftil_i\sqrt{N}(\daaa_i+\aaa_i) \right] \\&\phantom{=\sum}- J(\daaa_1\aaa_2+\daaa_2\aaa_1)\,,
	\eea
	where $\aaa_{1,2}$ are the two cavity modes, $\Delta$ is the detuning, $\Util$ is the on-site self-interaction strength, $\Ftil_i$ is the coherent drive, $J$ is the coupling amplitude
	and each cavity has single-photon loss $\Lhat_i=\sqrt{\kappa}\aaa_i$. 
	
	We consider the parameters $\Delta=-1.5\kappa$, $\Util=2\kappa$, $J=2.5\kappa$, and $\Ftil_1=-\Ftil_2=\Ftil$, which is the case studied in Ref.~\cite{delmonteMeasurementinducedPhaseTransitions2025}, where finite-$N$ simulations suggest the emergence of an entanglement \hl{criticality} coinciding with the dissipative phase transition. Fig.~\ref{fig:ept-bhd-qj-dynes} displays the steady-state solution as a function of the drive $\Ftil$ obtained with our analytical framework. Panel (a) shows the normalized population $|\alphatil_i|^2$ in the two cavities. In the stable solution of the steady state, the $\mathbb{Z}_2$ symmetry of the two spatial modes $\aaa_1\leftrightarrow-\aaa_2$ is spontaneously broken, resulting in a phase with different occupation in the two sites. Using this stable solution in the equations for the covariances, we obtain the stable solution for the steady-state entanglement entropy, as shown in panel (b) for the entanglement between the two spatial modes $S_E(\aaa_1,\aaa_2)$, and in panel (c) for the entanglement between the bonding and the antibonding modes $S_E(\aaa_+,\aaa_-)$ [with $\aaa_\pm\equiv (\aaa_1\pm\aaa_2)/\sqrt{2}$]~\footnote{The entanglement is computed using the Gaussian covariance matrix formalism in the same way as we did in Ref.~\cite{delmonteMeasurementinducedPhaseTransitions2025}}. As expected from the hierarchical structure of Eq.~\eqref{eq:det-boson-multi}, the non-analytical behavior of the population is also imprinted in the entanglement, where sharp transitions occur at the same mean-field critical points for all three unravelings (quantum jump, homodyne, and heterodyne) considered here. On the other hand, if we adopt the unstable solution for the mean-field $\alphatil_i$ [marked with a dotted line in panel (a)] with unbroken symmetry in the evaluation of the entanglement, we observe only a smooth behavior of the entanglement entropy as a function of the drive $\Ftil$ for all unravelings [marked with dotted lines in panels (b) and (c)] with no transition.
	In this regard, the non-analyticity in entanglement appears merely as a byproduct of the symmetry-breaking mean-field phase transition.

	\section{ Infinite-range spins}\label{sec:results-collective-spins}
	The theoretical framework developed for Bosons can be straightforwardly transposed to infinite-range spin systems. For concreteness, we consider the collective spin system of $N$ spin-$1/2$ particles subjected to collective drive (of amplitude $\Omega$) and dissipation (of rate $\kappa$) with the following Hamiltonian and jump operator~\cite{Passarelli},
	\bea\label{eq:collective-spin-ham-l}
	\Hhat = \Omega\Shat^x\,,\quad     \Lhat = \sqrt{\dfrac{\kappa}{S}}\Shat^-\,.
	\eea 
	Here, $\Shat^{x,y,z}\equiv \sum_{i=1}^N \hat{\sigma}_i^{x,y,z}/2$ and $\Shat^{\pm}=\Shat^x\pm\rmi\Shat^y$ are the collective spin operators composed from the Pauli operators $\sigma^\mu_i$ for the individual spins, and the total spin number of the system is $S=N/2$. \hl{In previous finite-$S$ numerical studies, this model (and some of its variants) has been shown to present a phase transition from a stationary phase to a time crystal phase accompanied by an entanglement criticality at the same critical point~\cite{delmonteMeasurementinducedPhaseTransitions2025,Passarelli,liMonitoredLongrangeInteracting2025a,Iemini}.}
	
	As derived in \hl{Appendix~\ref{app:collective-spin-eom}}, we follow the prescription detailed in Ref.~\cite{liMonitoredLongrangeInteracting2025a, holsteinFieldDependenceIntrinsic1940} to map the spins into a bosonic mode $\bbb$, which represents the quantum fluctuations around the collective spin, whose covariances $ u = \langle\delhat\delhat\rangle$ and $v = \langle\ddelhat\delhat\rangle$, with $\delhat=\bbb-\langle\bbb\rangle$, determine the entanglement present in the system (under the Gaussian assumption). As in the bosonic case, the resulting equations of motion for $u$ and $v$ [see Eqs.~\eqref{eq:spin-dtheta-dphi-det} and~\eqref{eq:spin-du-dv-det}] are deterministic in the $S\to\infty$ limit and are driven by the mean-field solution of the (classical) spin vector $m^\alpha\equiv\langle\Shat^\alpha\rangle/S$. 
	
	In the collective spin model in Eq.~\eqref{eq:collective-spin-ham-l}, when $\Omega<\kappa$, the system admits a stationary phase for the magnetization, characterized by the stable solution for the spin vector    $\mvec_\mathrm{ss} = (0,\Omega/\kappa,-\sqrt{1-\Omega^2/\kappa^2})$.
	From the equations for the fluctuations (see \hl{Appendix~\ref{app:collective-spin-eom}}), one can also obtain an analytical solution to the steady-state covariances,
	\bea\label{eq:spin-uvss-ana}
	u_\mathrm{ss} = \dfrac{\Omega^2}{4\kappa\sqrt{\kappa^2-\Omega^2}}\,,\quad 
	v_\mathrm{ss} = -\dfrac{-\kappa+\frac{\Omega^2}{2\kappa}+\sqrt{\kappa^2-\Omega^2}}{2\sqrt{\kappa^2-\Omega^2}}\,,
	\eea
	that is shared by all the different unravelings.
	Since the half-system entanglement $S_E$ is a function of $v$~\cite{leroseOriginSlowGrowth2020} \hl{along a single trajectory},
	\bea\label{eq:collective-ent}
	S_E(v) = \sqrt{1+v}\coth^{-1}\left(\sqrt{1+v}\right)+\dfrac{1}{2}\log\left(\dfrac{v}{4}\right)\,.
	\eea
	we obtain the steady-state entanglement upon inserting $v_\mathrm{ss}$ into the expression above \hl{(see Appendix~\ref{app:collective-spin-benchmark} for a benchmark of the entanglement dynamics against finite-$S$ trajectories)}.
	
\begin{figure}[t]
		\centering
		\includegraphics[width=0.9\linewidth]{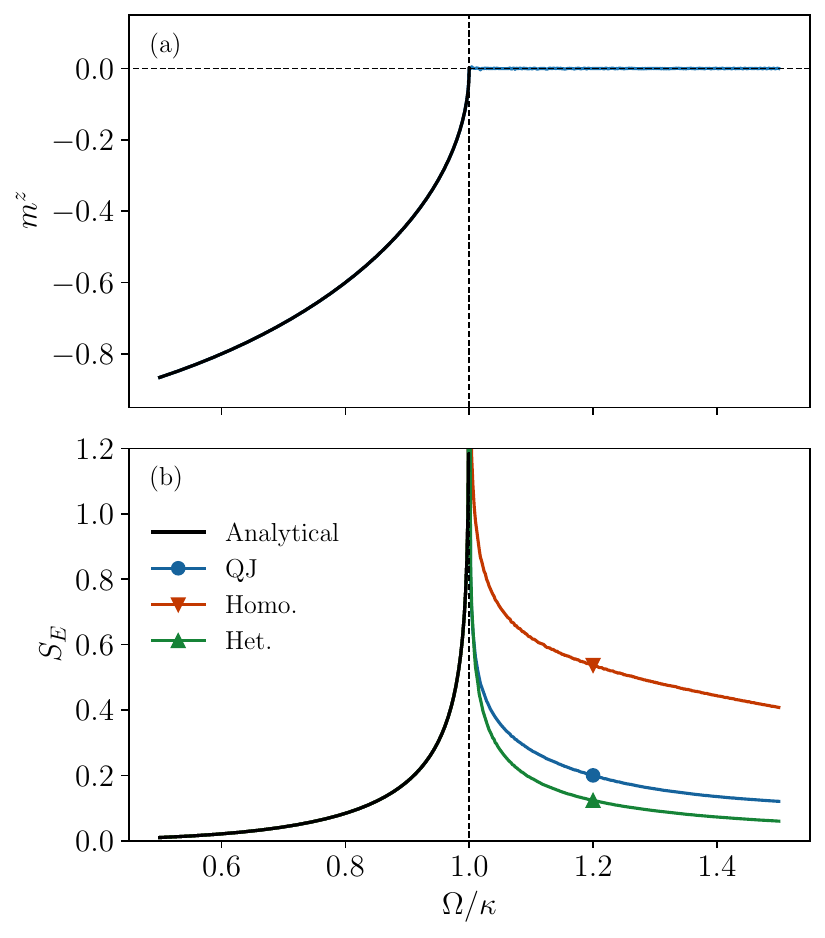}
		\caption{Long-time averaged solution of the deterministic equations in the thermodynamic limit of the collective spin model for (a) the magnetization $m^z$ and (b) the half-system entanglement entropy $S_E$ with different unravelings [quantum jump (QJ), homodyne (Homo.) and heterodyne (Het.), see legend]. In both panels, the solid line for $\Omega<\kappa$ represents the analytical solution (for $m^z$ and $S_E$ respectively) in the stationary phase. The vertical dashed line in both panels marks the mean-field critical point at $\Omega=\kappa$.}
		\label{fig:sz_ent_ss}
	\end{figure}

	The analytical solution for the steady-state magnetization $m^z$ and for the half-system entanglement entropy are shown in Fig.~\ref{fig:sz_ent_ss} for $\Omega<\kappa$. When the drive approaches the mean-field critical point $\Omega\to\kappa_-$, the steady-state entanglement also diverges [since $\lim_{\Omega\to\kappa_-}(v_\mathrm{ss})=\infty$ from Eq.~\eqref{eq:spin-uvss-ana}]~\footnote{Note that for any $\Omega<\kappa$, the solution for the covariances remains finite, which does not violate our assumption that $u\sim v\sim\Ocal(1)$. We can conclude the divergence by sending $\Omega$ arbitrarily close to the critical point, where the covariances also grow arbitrarily large.}. Similar to the Bosonic case discussed earlier, we have analytically established the coincidence of the entanglement criticality with the dissipative phase transition at the mean-field critical point.

	\begin{figure}[t]
		\centering
		\includegraphics[width=0.9\linewidth]{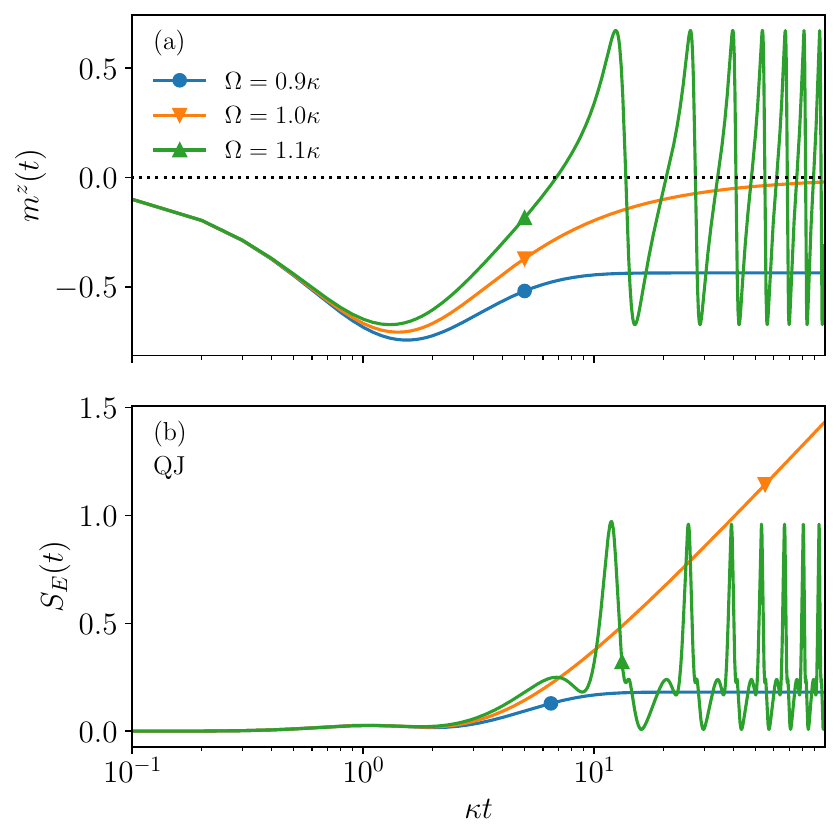}
		\caption{Time evolution (in linear-log scale) of the (a) magnetization $m^z$ and (b) the half-system entanglement entropy $S_E$ [with quantum-jump (QJ) unraveling] of the collective spin model in the thermodynamic limit with different values of the drive $\Omega$ (see legend).}
		\label{fig:sz_ent_qj_dyn}
	\end{figure}
	
	When $\Omega>\kappa$, this stationary solution ceases to exist and the mean-field theory predicts a time-crystal phase with permanent oscillating magnetization~\cite{Iemini}. Solving the dynamics numerically (with an initial state where all spins are aligned in $+x$ direction), we obtain the time evolution of both the magnetization $m^z$ and the half-system entanglement entropy $S_E$ (with quantum-jump unraveling) for different values of the drive $\Omega$, as shown in Fig.~\ref{fig:sz_ent_qj_dyn}. At $\Omega=0.9\kappa$ in the stationary phase, both the magnetization and the entanglement reach their steady-state value predicted by our analytical solution. At critical driving $\Omega=\kappa$, the magnetization relaxes asymptotically towards $m^z=0$ (with a diverging relaxation time), whereas the entanglement diverges with time, displaying a logarithmic growth, suggesting the emergence of an entanglement criticality. In the time-crystal phase with $\Omega=1.1\kappa$, both the magnetization and the entanglement show permanent oscillations. 
	
	Finally, we show in Fig.~\ref{fig:sz_ent_ss} the long-time averaged magnetization $m^z$ and entanglement $S_E$ as a function of the drive $\Omega$ and the unraveling (quantum jump, homodyne and heterodyne) in the thermodynamic limit. For both quantities, the values in the stationary phase ($\Omega<\kappa$) are obtained with the analytical solution; in the time-crystal phase ($\Omega>\kappa$), we performed a temporal average from $\kappa t=3000$ to $\kappa t=4000$. For all three unravelings considered, our results show an entanglement \hl{criticality} coinciding with the stationary-to-time-crystal phase transition, in agreement with previous studies~\cite{delmonteMeasurementinducedPhaseTransitions2025,Passarelli,liMonitoredLongrangeInteracting2025a}.

	\section{ Conclusions}\label{sec:conclu}
	We have studied entanglement dynamics in the thermodynamic limit of monitored infinite-range interacting systems. Our results, exploiting the semiclassical nature of infinite-range systems, reveal several features that can be highlighted as follows,
	\begin{itemize}
		\item Along single trajectories, the normalized observables follow deterministic dynamics in the thermodynamic limit given by the standard mean-field theory, which drives the (also deterministic)  dynamics of quantum fluctuations and in particular, the entanglement.
		\item When the mean-field observables exhibit non-analytical behavior, this also drives the appearance of an entanglement \hl{criticality} due to the one-way dependence of the latter on the former.
	\end{itemize}
	
	This mean-field picture for monitored entanglement dynamics complements our previous finite-size numerical results~\cite{Passarelli,delmonteMeasurementinducedPhaseTransitions2025,liMonitoredLongrangeInteracting2025a}. The unified treatment for different unravelings makes it a convenient toolbox for studying monitored dynamics in the semiclassical regime, that can be readily applied to various infinite-range models beyond those considered in this work.

    \section*{Data availability}
	\hl{The data generated in this study has been deposited in the Zenodo public folder~\cite{li_2025_15739056}.}
	
	\section*{ Acknowledgments}
	We would like to thank Xhek Turkeshi,  Gianluca Passarelli, Cristiano Ciuti, Jacopo Tosca and Valentin Heyraud for helpful discussions. This work was supported by PNRR MUR project PE0000023- NQSTI, by the European Union (ERC, RAVE, 101053159). Views and opinions expressed are however those of the author(s) only and do not necessarily reflect those of the European Union or the European Research Council. Neither the European  Union nor the granting authority can be held responsible for them.

	\appendix

	\section{ Semiclassical limit of the quantum-jump unravelling}\label{app:qj-semi}
	We consider the quantum-jump stochastic \schr{} equation~\cite{Wiseman},
	\bea\label{eq:sse-jump}
	\d|\psi\rangle &=  \d t\left[ - \rmi \Hhat+\sum_i\left( \langle\Lhat^\dagger_i\Lhat_i\rangle - \Lhat^\dagger_i\Lhat_i \right)\big/2   \right]|\psi\rangle \\ 
	&+ \sum_i \d M_i(t)\left( \Lhat_i\Big/\sqrt{\langle\Lhat^\dagger_i\Lhat_i\rangle}-\idhat \right)|\psi\rangle \,,
	\eea
	where $M_i(t)$ denotes the number of jumps occurred up to time $t$ associated with the jump operator $\Lhat_i$. The point-processes $\d M_i(t)$ satisfy 
	\bea\label{eq:dm-jump}
	\d M_i \d M_j = \d M_i \delta_{ij}\,,\quad \mathbb{E}[\d M_i] = \langle\Lhat_i^\dagger\Lhat_i\rangle\d t\,.
	\eea
	Within our semiclassical assumptions, we consider the extensive case where $\langle\Lhat_i\rangle\sim \Theta(\sqrt{N})$. Setting the small parameter $\epsilon\sim1/\sqrt{N}$, we consider a coarse-grained time scale $\Delta t\sim \epsilon\sqrt{\epsilon}$ that is small enough for the relevant system dynamics but long enough for the mean number of jumps in each channel $\mu_i\simeq |\langle\Lhat_i\rangle|^2\Delta t\sim 1/\sqrt{\epsilon}$ to be macroscopic. 
	The coherent approximation~\eqref{eq:coherent-approx} immediately implies that the number of jumps $\Delta M_i$ occurring in channel $i$ within time $\Delta t$ follows a Poissonian distribution with both mean and variance equal to $\mu_i$. Since $\mu_i\sim1/\sqrt{\epsilon}$ diverges with $N$, the Poissonian can be approximated by a Normal distribution of the same mean and variance~\cite{wisemanQuantumTheoryFieldquadrature1993}. We therefore treat the number $\Delta M_i$ as a random variable defined by
	\bea\label{eq:m-random}
	\Delta M_i = |\langle\Lhat_i\rangle|^2\Delta t + |\langle\Lhat_i\rangle|\Delta W_i\,,
	\eea
	where the $\Delta W_i$ are independent Weiner increments with zero mean and $\mathbb{E}[\Delta W_i^2]=\Delta t$. Replacing the increment $\d M_i$ in Eq.~\eqref{eq:sse-jump} with the coarse-grained version~\eqref{eq:m-random} and taking the limit $\Delta t \to \d t$, $\Delta W \to \d W$, we obtain the stochastic \schr{} equation in the large $N$ limit, which can be compactly written for the density matrix $\rhohat \equiv \ketbra{\psi}$ (using It\^{o} differentiation rules)~\footnote{Note that we also assume the factorization $\langle\Lhat^\dagger_i\Lhat_i\rangle=|\langle\Lhat_i\rangle|^2$ in Eq.~\eqref{eq:sse-jump}, as allowed by the coherent approximation~\eqref{eq:coherent-approx}.},
	\bea
	\d\rhohat = \d t
	\lcal\rhohat + \sum_i[ \d Z^*_i(\Lhat_i-\langle\Lhat_i\rangle) \rhohat+\hc]\,,
	\eea
	with the state-dependent complex noise $\d Z_i\equiv \langle\Lhat_i\rangle\d W_i/|\langle\Lhat_i\rangle|$. This gives Eq.~\eqref{eq:mf-coarse-general} in the main text, which is of the form of a general-dyne unraveling.

	\subsection{\hl{An explicite derivation for a single-mode problem}}
	We provide here an explicit derivation of the coarse-grained quantum-jump stochastic master equation, for a single-mode problem with $\Lhat = \sqrt{\kappa}\aaa$. Denoting
	\bea
	\Ucal(t)\rhohat\equiv \rme^{-\rmi\Hhat_\nh\hl{t}}\rhohat \rme^{\rmi\Hhat_\nh^\dagger\hl{t}}\,,\quad \Jcal\rhohat \equiv \Lhat\rhohat\Lhat^\dagger\,,
	\eea
	where the non-Hermitian Hamiltonian is defined as
	\bea
	\Hhat_\nh \equiv \Hhat - \dfrac{\rmi}{2}\Lhat^\dagger\Lhat\,,
	\eea
	the probability for $m$ jumps to occur within time $\Delta t$ starting from the state $\rhohat$ is~\cite{wisemanQuantumTheoryFieldquadrature1993}
	\bea
	\mathbb{P}(m;\Delta t&) = \Tr\Bigg[\int_0^{\Delta t}\d t_m\int_0^{t_m}\d t_{m-1}\cdots\int_0^{t_2}\d t_1\\ 
	&\Ucal(\Delta t-t_m)\Jcal\Ucal(t_m-t_{m-1})\cdots\Jcal\Ucal(t_1)\rhohat \Bigg]\,.
	\eea
	The semiclassical approximations made in the main text allow $\Ucal$ and $\Jcal$ to commute at leading order, making the nested integral trivial to evaluate,
	\bea
	\mathbb{P}(m;\Delta t)\simeq\dfrac{(\Delta t)^m}{m!}\Tr\left[ \Ucal(\Delta t)\Jcal^m\rhohat \right]\simeq \rme^{-\mu} \dfrac{\mu^m}{m!}\,,
	\eea
	where we remind that $\mu\equiv|\langle\Lhat\rangle|^2\Delta t=|\alpha|^2\kappa\Delta t$. This is a Poissonian distribution with both mean and variance equal to $\mu$. This is not surprising as we are assuming the state to be coherent at leading order, which has Poissonian counting statistics. Approximating the number of jumps $m$ with the following Gaussian random number (as justified in the main text),
	\bea\label{eq:m-random-sm}
	m = |\alpha|^2\kappa\Delta t + |\alpha|\sqrt{\kappa}\Delta W\,,
	\eea
	we find the state conditioned by $m$ jumps [using the expansion $\aaa=\alpha(\idhat+\delhat/\alpha)$],
	\bea
	|\tilde{\psi}(\Delta t)\rangle &\propto \rme^{-\rmi \Hhat_\nh\Delta t}\aaa^m|\psi\rangle\\
	& \propto \rme^{-\rmi\Hhat_\nh\Delta t}\left(\idhat + \dfrac{\delhat}{\alpha}\right)^m|\psi\rangle\\
	&\simeq \left( \idhat - \rmi\Hhat_\nh\Delta t \right)\left( \idhat + m\dfrac{\delhat}{\alpha} \right)|\psi\rangle \\
	&\simeq |\psi\rangle + \left[ m\left( \dfrac{\aaa}{\alpha} - \idhat \right) -\rmi\Hhat_\nh\Delta t\right]|\psi\rangle\,.
	\eea
	After normalizing the state, we get the increment
	\bea
	\Delta^{(m)}|\psi\rangle =&~  m\left( \dfrac{\aaa}{\alpha} -\idhat\right)|\psi\rangle \\
	&+\Delta t\left[\kappa
	\left( \dfrac{|\alpha|^2}{2}-\dfrac{\daaa\aaa}{2} \right)-\rmi\Hhat\right]|\psi\rangle\,,
	\eea
	which is of the same form as the original stochastic \schr{} equation (A1) in the main text with the point increment $\d M$ replaced by the coarse-grained random number $m$. Taking the limit $\Delta t \to \d t$, $\Delta W \to \d W$ and inserting the expression for $m$ from Eq.~\eqref{eq:m-random-sm}, we arrive at the stochastic \schr{} equation for the normalized pure state in the large $N$ limit,
	\bea
	\d|\psi\rangle =&~ \d t \left[  |\alpha|^2\kappa \left( \dfrac{\aaa}{\alpha} - \idhat \right) + \dfrac{\kappa|\alpha|^2}{2} - \dfrac{\kappa\daaa\aaa}{2} - \rmi\Hhat \right]|\psi\rangle \\
	&+ |\alpha|\sqrt{\kappa}\d W\left(\dfrac{\aaa}{\alpha}-\idhat\right)|\psi\rangle\,.
	\eea
	Equivalently, in terms of the density matrix $\rhohat \equiv \ketbra{\psi}$, the above equation can be written as
	\bea
	\d \rhohat = \d t\Lcal\rhohat + |\alpha|\sqrt{\kappa}\d W \left( \dfrac{\aaa-\alpha}{\alpha}\rhohat + \rhohat\dfrac{\daaa-\calpha}{\calpha}  \right)\,,
	\eea
	which is essentially Eq.~\eqref{eq:mf-coarse-general}  of the main text.

	\section{ Deterministic dynamics in the thermodynamic limit}\label{app:det-limit}
	The evolution of the first moments is obtained by
	the adjoint stochastic master equation
	\bea
	\d\alpha_i = \d t \langle\Lcal^\ddagger\aaa_i\rangle+\langle\d\hcal^\ddagger\aaa_i\rangle\,,
	\eea
	where $\acal^\ddagger$ denotes the adjoint of the superoperator $\acal$ defined via $\tr\{(\acal^\ddagger\ohat)\rhohat\}\equiv\tr\{\ohat(\acal\rhohat)\}$. As we require the model to be extensive, i.e., $\langle\Hhat\rangle\sim \Theta(N)$ and $\langle\Lhat_i\rangle\sim\Theta(\sqrt{N})$, the Hamiltonian $\Hhat$ should consist of terms of the form $N^{1-\frac{m+n}{2}}\aaa^{\dagger m}\aaa^n$, and $\Lhat$ takes the form $N^{\frac{1}{2}-\frac{m+n}{2}}\aaa^{\dagger m}\aaa^n$ [due to the coherent approximation~\eqref{eq:coherent-approx}].
	Exploiting the bosonic commutation relations and the Gaussian approximation, it then follows that $\langle\Lcal^\ddagger\aaa_i\rangle\sim\Theta(\sqrt{N})$, which is dominant over the noise term $\langle\d\hcal^\ddagger\aaa_i\rangle\sim\Ocal(1)$. Defining the normalized first moment $\alphatil_i\equiv \alpha_i/\sqrt{N}$, we immediately recover the standard deterministic mean-field equation in the $N\to\infty$ limit, which involves only $\alphatil_i$ but not the covariances $u_{ij}$ or $v_{ij}$.
	
	On the other hand, the quantum fluctuations encoded in the covariances exhibit nontrivial dynamics dependent on the first moment and, in particular, on the monitoring. Applying the It\^{o} differentiation rule to the time-dependent operator $\delhat_i\delhat_j$, we obtain
	\bea\label{eq:du-kerr-raw}
	\d u_{ij} &= \d t\langle\lcal^\ddagger(\delhat_i\delhat_j)\rangle + {\langle\d\hcal^\ddagger(\delhat_i\delhat_j)\rangle} - \d \alpha_i \d \alpha_j\,,
	\eea
	where both $\langle\lcal^\ddagger(\delhat_i\delhat_j)\rangle$ and $\d \alpha_i \d \alpha_j$ are of order $1$ while the stochastic term is of order $1/\sqrt{N}$. Taking the $N\to\infty$ limit, the resulting equation is therefore deterministic, and the same holds true for $v_{ij}$.

	\begin{figure*}[ht]
		\centering
		\includegraphics[width=\linewidth]{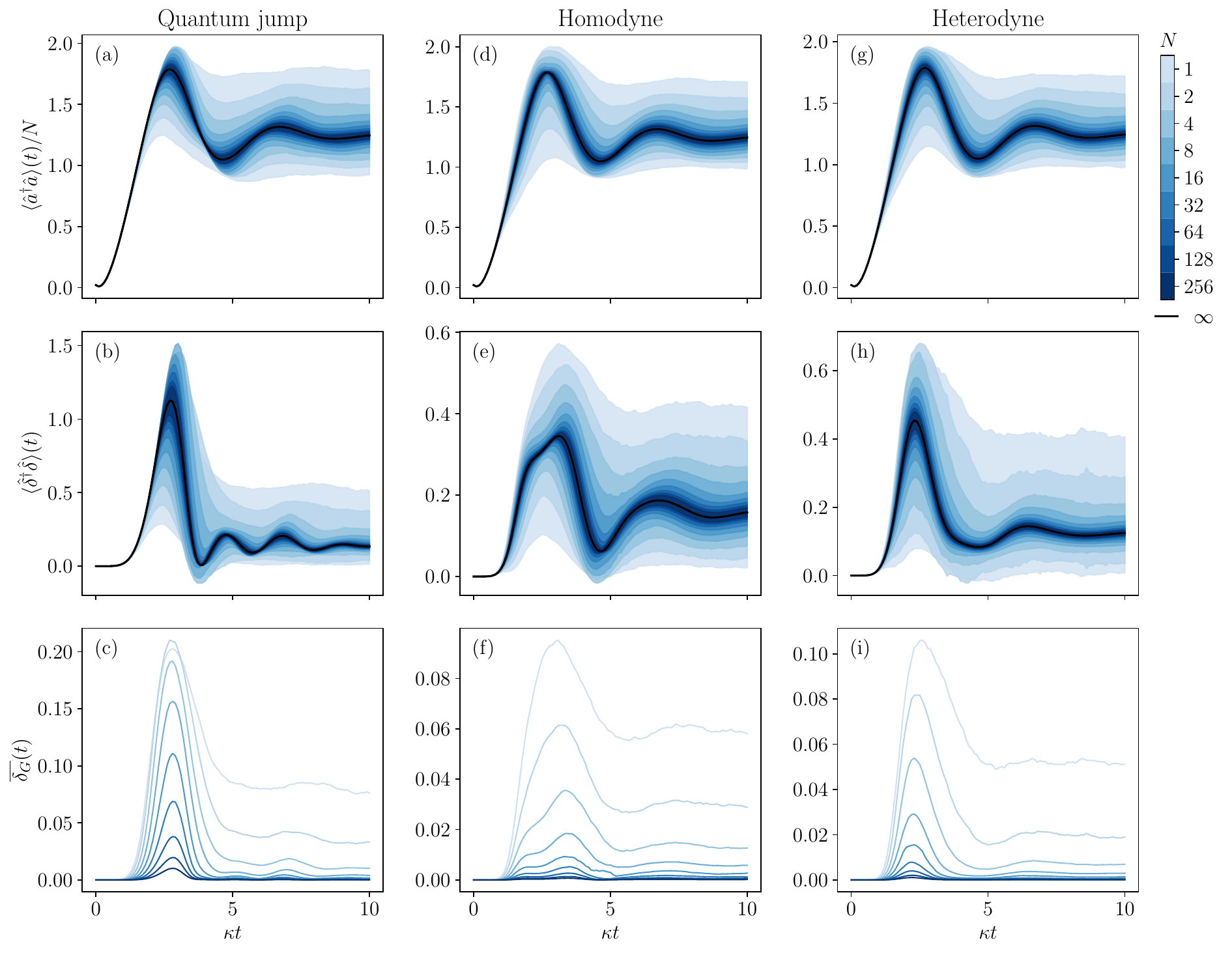}
		\caption{Benchmark of the deterministic equations for a single-mode Kerr cavity against numerically exact trajectories for finite $N$ (see legend) and different unravelings: the first column [panels (a)-(c)] for quantum jump, the second column [panels (d)-(f)] for homodyne and the third column [panels (g)-(i)] for heterodyne unraveling. The first row [panels (a), (d) and (g)] shows the dynamics of the normalized Bosonic population $\langle\daaa\aaa\rangle/N$ in the cavity, and the second row [panels (b), (e) and (h)] shows the time evolution of the covariance $v=\langle\ddelhat\delhat\rangle$. The shaded region corresponds to one standard deviation around the trajectory-average values, and darker color represents larger values of $N$. In all three unravelings considered, we observe for both quantities the reduction of statistical fluctuations with increasing $N$ and the asymptotic convergence of the distribution towards the result obtained by the deterministic equation (marked by solid black lines). The third row [panels (c), (f) and (i)] shows the trajectory-average value of the non-Gaussianity measure, $\overline{\delta_G}$, for the finite-$N$ trajectories. The collapse of the lines towards $0$ with increasing $N$ suggests the asymptotic Gaussianity of the single-trajectory state along the monitored stochastic evolution.  Parameters: $\Ftil=\Util=\kappa$ and $\Delta=0.5\kappa$. Around 3000 trajectories are performed in each finite-$N$ simulation. The homodyne and heterodyne trajectories are simulated with a fixed time step $\kappa\d t=10^{-6}$.}
		\label{fig:single-kerr-bench}
	\end{figure*}
	
	\section{Benchmark on a single-mode bosonic system}\label{app:benchmark-single-kerr}
	
	To visualize the emergence of the deterministic dynamics in the thermodynamic limit, we provide in this section a benchmark based on a single-mode Kerr cavity model against numerically exact quantum trajectories simulated at finite $N$. 
	The considered Hamiltonian is
	\bea
	\Hhat = -\Delta
	\daaa\aaa+\dfrac{\Util}{2 N}\daaas\aaas+\Ftil\sqrt{N}(\daaa+\aaa)\,
	\eea
	and the jump operator is $\Lhat=\sqrt{\kappa}\aaa$. The deterministic equations (7) in the main text give the following,
	\bea\label{eq:single-kerr-det}
	\d \alphatil = \d t\Big[ &\Big(\rmi\Delta-\dfrac{\kappa}{2}\Big)\alphatil - \rmi\Util|\alphatil|^2\alphatil-\rmi\Ftil \Big]\,,\\
	\d u = \d t\Big\{&(2\rmi\Delta-\kappa)u -\rmi \Util ( \alphatil^2 + 2\alphatil^2 v + 4|\alphatil|^2 u ) \\ 
	&- \kappa \left( 2uv + \Upsilon^*u^2+\Upsilon v^2 \right)\Big\}\,,\\
	\d v = \d t\Big\{ &2\Util\Im[\alphatil^2u^*]-\kappa v\\
	&- \kappa\left(2\Re[\Upsilon u^* v]+|u|^2+ v^2 \right) \Big\}\,,
	\eea
	and we remind that $\Upsilon$ is the unraveling-dependent factor defined in Eq. (6) of the main text.
	Fig.~\ref{fig:single-kerr-bench} shows the temporal dynamics of the system obtained from three different unravelings (quantum jump, homodyne, and heterodyne) with $\Ftil=\Util=\kappa$ and $\Delta=0.5\kappa$. The initial state is chosen to be a coherent state $\ket{\alpha}$ with $\alpha/\sqrt{N}=0.1+0.1\rmi$. We first consider the normalized Bosonic population $\langle\daaa\aaa\rangle/N$. As shown in panels (a), (d), and (g), the statistical fluctuations of quantum trajectories are represented as shaded regions in the graph, which is centered at the trajectory-average value with a radius equal to the standard deviation (among single-trajectory values). As $N$ increases, the distribution of trajectories narrows down and converges towards the deterministic solution. This validates the standard mean-field equation for the normalized first moment along single trajectories. 
	
	The same benchmark is performed in panels (b), (e), and (h), but for the nonlinear quantity $v=\langle\ddelhat\delhat\rangle$. Again, we observe the asymptotical convergence of the trajectory distribution towards the deterministic solution given by Eq.~\eqref{eq:single-kerr-det}. Note that the different dynamics shown in the three panels are due to the effect of different unravelings (i.e., measurement back-action), which is correctly captured by our deterministic equations. Importantly, the entanglement present in this collective model is uniquely determined by $v$ [cf. Eq. (11) of the main text]. This demonstrates the emergence of the deterministic entanglement dynamics in the infinite-range Bosonic model as predicted by our theory.
	
	Finally, let us verify the validity of the Gaussian approximation along single trajectories. Due to the interacting nature of the system, we expect the state to be non-Gaussian for finite $N$, and we quantify the non-Gaussianity of the state using the measure proposed in Ref.~\cite{genoniMeasureNonGaussianCharacter2007}. For a generic state $\rhohat$ with first moment $\alpha=\tr[\rhohat\aaa]$ and covariances $u=\tr[\rhohat\delhat\delhat]$, $v=\tr[\rhohat\ddelhat\delhat]$, we can always find a reference Gaussian state $\rhohat_\gcal(\alpha,u,v)$ with the same first and second moments. The non-Gaussianity measure is then defined as
	\bea
	\delta_G \equiv \dfrac{\tr[(\rhohat-\rhohat_\gcal)^2]}{2\tr[\rhohat^2]}\,,
	\eea
	which is the normalized Hilbert-Schmidt distance between the two density matrices. This quantity equals zero if and only if the state $\rhohat$ is Gaussian and is otherwise strictly positive. In Fig.~\ref{fig:single-kerr-bench} panels (c), (f), and (i), we show the trajectory-averaged non-Gaussianity $\overline{\delta_G}$ along the time evolution. For all three unravelings considered above, the non-Gaussianity asymptotically decreases to zero as $N$ increases, suggesting the emergent Gaussianity of the state in the thermodynamic limit. Interestingly, the finite-$N$ results also display a systematically higher non-Gaussianity in the quantum-jump trajectories compared to the state-diffusion unravelings. This implies that the Gaussian trajectory approximation might yield physically more accurate results with the diffusion-type unravelings in simulating finite-$N$ quantum trajectories, as compared to the (numerically more efficient) quantum-jump unraveling. Nevertheless, since our theoretical treatment directly targets the thermodynamic limit, we can safely assume the Gaussianity of the state.
	
	\section{Equations of motion for the collective spin model in the thermodynamic limit}\label{app:collective-spin-eom}
	
	In this section, we derive the expressions for the equations of motion of the collective spin system. We consider a slightly more general form for the Hamiltonian and jump operator,
	\bea\label{eq:collective-spin-model-def}
	\Hhat = \Omega\sum_\alpha c^F_\alpha\Shat^\alpha\,,\quad \Lhat = \sqrt{\dfrac{\kappa}{S}}\sum_\alpha c^D_\alpha\Shat^\alpha\,,
	\eea
	and the model defined in the main text corresponds to the choice   $\vec{c}^F = (1,0,0)$ and  $\vec{c}^D = (1,-\rmi,0)$. 
	
	We follow the same procedure as detailed in Ref.~\cite{liMonitoredLongrangeInteracting2025a} to bosonize the spin operators via the lowest-order Holstein-Primakoff transformation~\cite{holsteinFieldDependenceIntrinsic1940},
	\bea\label{eq:bosonization-n}
	\Shat^{\ztil} &= S - \dbbb\bbb\,,\\
	\Shat^{\xtil} &\simeq  \sqrt{\dfrac{S}{2}}(\dbbb + \bbb)\,,\\
	\Shat^{\ytil} &\simeq  \rmi\sqrt{\dfrac{S}{2}}(\dbbb - \bbb)\,,
	\eea
	with $[\bbb,\dbbb]=1$, in the rotated frame defined by $\Uhat(\theta,\phi) = \rme^{-\rmi\phi\Shat^z}\rme^{-\rmi\theta\Shat^y}$ and
	\bea
	\Shat^\alphatil = \Uhat(\theta,\phi)\Shat^\alpha\Uhat^\dagger(\theta,\phi)=\sum_\beta G_{\alphatil\beta}\Shat^\beta\,,
	\eea
	with the following matrix elements,
	\bea
	G &\equiv \pmx{
		\cos\theta\cos\phi & \cos\theta \sin\phi  &  -\sin\theta\\
		-\sin\phi & \cos\phi & 0  \\
		\sin\theta\cos\phi & \sin\theta\sin\phi  & \cos\theta
	}\,.
	\eea
	A generic spin operator of the form $\Khat = \sum_\beta c_\beta\Shat^\beta$ can then be written in the bosonized picture as
	\bea\label{eq:spin-op-bosonic}
	\Khat &= S\ccal_\ztil + \sqrt{\dfrac{S}{2}}( \ecal\dbbb + \fcal^*\bbb ) - \ccal_\ztil\dbbb\bbb\,,
	\eea
	where we have defined 
	\bea\label{eq:cefcal}
	\ccal_\alphatil&\equiv\sum_\beta c_\beta G_{\alphatil\beta}\,,\\\ecal&\equiv \ccal_\xtil + \rmi \ccal_\ytil\,,\\ \fcal&\equiv\ccal^{*}_\xtil + \rmi \ccal^{*}_\ytil\,.
	\eea
	The dynamics can be studied in the rotating frame by introducing the inertial Hamiltonian
	\bea\label{eq:ham-rf}
	\Hhat_\rf &= -\rmi \dfrac{\d \uhat}{\d t}\uhat^\dagger\\&= \sin\theta\dfrac{\d \phi}{\d t}\Shat^\xtil - \dfrac{\d \theta}{\d t}\Shat^\ytil - \cos\theta\dfrac{\d\phi}{\d t}\Shat^\ztil\,.
	\eea
	The equation of motion for the first moment $\beta\equiv\langle\bbb\rangle$ is then given by,
	\bea
	\d\beta =&~ \rmi\d t\langle[\Hhat_\rf\,,\bbb]\rangle + \d t\langle\Lcal^\ddagger\bbb\rangle+ \langle\d\hcal^\ddagger\bbb\rangle\,,
	\eea
	where $\lcal^\ddagger$ is the adjoint bare Liouvillian (without $\Hhat_\rf$).
	The action of the inertial Hamiltonian is explicitly evaluated to be
	\bea
	\rmi[\Hhat_\rf\,,\bbb]=\rmi\sqrt{\dfrac{S}{2}}\sin\theta\dfrac{\d\phi}{\d t} - \sqrt{\dfrac{S}{2}}\dfrac{\d\theta}{\d t}-\rmi\cos\theta\dfrac{\d\phi}{\d t}\bbb\,.
	\eea
	Finally, we impose the alignment of the rotating frame's $\ztil$ axis with the collective magnetization (i.e. $\langle\bbb\rangle\equiv0$) and using the fact that the noise term $\langle\d\hcal^\ddagger\bbb\rangle$ is of order $\Ocal(1)\d Z$ while the leading order of both $\langle\Lcal^\ddagger\bbb\rangle$ and $\langle[\Hhat_\rf\,,\bbb]\rangle$ is $\Theta(\sqrt{S})$,  we obtain, in the $S\to\infty$ limit, the equations of motion for the angles $(\theta,\phi)$,
	\bea\label{eq:spin-dtheta-dphi-det}
	\sin\theta\dfrac{\d\phi}{\d t} &= \left\{ \sqrt{\dfrac{2}{S}}\Im\langle\Lcal^\ddagger\bbb\rangle \right\}_0 \,,\\
	\dfrac{\d \theta}{\d t} &= \left\{\sqrt{\dfrac{2}{S}}\Re\langle\Lcal^\ddagger\bbb\rangle\right\}_0\,.
	\eea
	They result in the standard mean-field equations for the classical spin vector $(\theta,\phi)$ in spherical coordinates.
	The deterministic equations for the second moments also follow naturally by a similar argument,
	
	\bea\label{eq:spin-du-dv-det}
	\d u &= -2\rmi u \cos\theta\dfrac{\d\phi}{\d t}\d t + \left\{ \d t\langle\Lcal^\ddagger(\delhat\delhat)\rangle - \langle\d\hcal^\ddagger\bbb\rangle^2\right\}_0\,,\\ \d v&=\left\{\d t\langle\lcal^\ddagger(\delhat^\dagger\delhat)\rangle-\left\lvert\langle\d\hcal^\ddagger\bbb\rangle\right\rvert^2\right\}_0\,,
	\eea
	
	Below, we provide the explicit expressions of these equations for the model defined in Eq.~\eqref{eq:collective-spin-model-def}. The right-hand side of Eq.~\eqref{eq:spin-dtheta-dphi-det} is given by 
	\bea
	\left\{\sqrt{\dfrac{2}{S}}\langle\Lcal^\ddagger\bbb\rangle\right\}_0\equiv-  \rmi \Omega\ecalf - \dfrac{\kappa}{2} \(\fcald\czd-\ecald\czdc\)
	\,.
	\eea
	The equations~\eqref{eq:spin-du-dv-det} for $u$ and $v$ become,
	\bea
	\dfrac{\d u}{\d t} =&{}~ 2\rmi\[\Omega  \ccal_\ztil^F -\cos\theta\dfrac{\d\phi}{\d t} \] u \\
	& -\dfrac{\kappa}{2}\left\{  \ecald\fcald+\(\abss{\fcald}-\abss{\ecald}\)u  \right\} \\
	&- \kappa \[ \ecald(v+1)+\fcaldc u \]\(\ecaldc u + \fcald v\)\\
	&- \dfrac{\kappa}{2}\Upsilon^*\[ \ecald(v+1)+\fcaldc u \]^2\\
	&- \dfrac{\kappa}{2}\Upsilon\(\ecaldc u + \fcald v\)^2\,,\\
	\dfrac{\d v}{\d t} =&{}
	-\dfrac{\kappa}{2}\[ \( \abss{\fcald}-\abss{\ecald} \)v-\abss{\ecald} \] \\
	&-\kappa\Re\left\{ \Upsilon \[ \ecaldc(v+1)+\fcald u^* \]\(\ecaldc u + \fcald v\) \right\}\\
	&- \dfrac{\kappa}{2}\left\lvert \ecald(v+1)+\fcaldc u \right\rvert^2\\
	&- \dfrac{\kappa}{2}\left\lvert\ecaldc u + \fcald v\right\rvert^2\,, 
	\eea
	where we adopt the definition of the angle-dependent coefficients in Eq.~\eqref{eq:cefcal}.

	\begin{figure}[t]
		\centering
		\includegraphics[width=\linewidth]{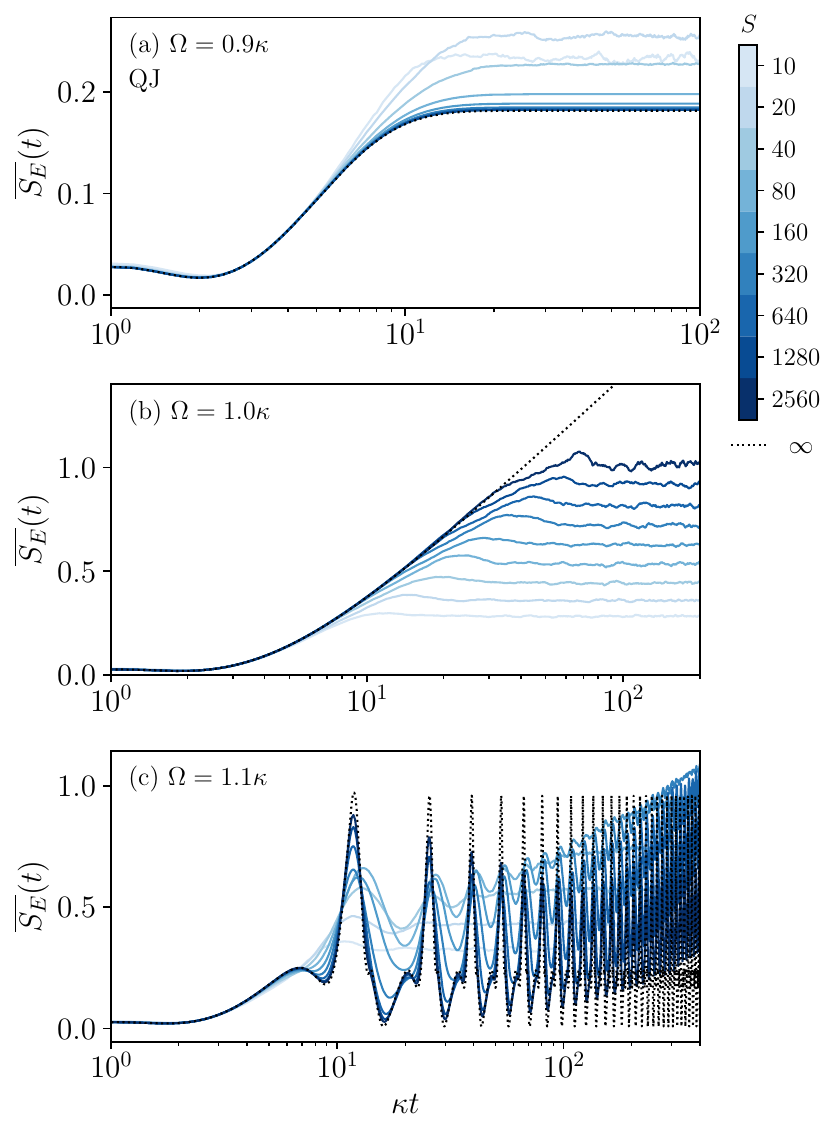}
		\caption{Finite-size results for the entanglement dynamics of the collective spin model with quantum-jump (QJ) unraveling for the three drive values considered in Fig. (3) of the main text: (a) $\Omega=0.9\kappa$, (b) $\Omega=1.0\kappa$ and (c) $\Omega=1.1\kappa$. The numerically exact results are denoted by solid lines with darker color representing larger $S$ (see legend). The solution in the thermodynamic limit is marked with dotted lines in comparison. Up to 5000 trajectories are performed in each finite-$S$ simulation.}
		\label{fig:sz_ent_dyn_exact_qj}
	\end{figure}
	
	\section{Benchmark for the collective spin system}\label{app:collective-spin-benchmark}
	
	We provide here a benchmark for the temporal dynamics predicted by the deterministic equations for the collective spin system. Fig.~\ref{fig:sz_ent_dyn_exact_qj} shows the time dynamics of the trajectory-averaged entanglement entropy $\overline{S_E}$ obtained from quantum-jump trajectories simulated for finite $S$ together with the solution of the deterministic equations for $S=\infty$. In the stationary phase with $\Omega=0.9\kappa$ [panel (a)], the results at large $S$ are practically indistinguishable from the $S=\infty$ solution. At criticality $\Omega=\kappa$ [panel (b)], while the thermodynamic limit gives an indefinite growth of entanglement, the results for finite $S$ always remain finite, which saturate towards a certain steady-state value that increases with $S$. Similarly, in the time-crystal phase $\Omega=1.1\kappa$ [panel (c)], the finite-$S$ entanglement deviates from the $S=\infty$ solution at long times and saturates at a steady-state value. This suggests that the $t\to\infty$ limit does not commute with $N\to\infty$. At any given finite time, however, both panels (b) and (c) show that the agreement between the finite-$S$ result and the $S=\infty$ result is improved increasing $S$. This suggests the asymptotic accuracy of our deterministic equations in evaluating quantities \textit{at finite times} in the thermodynamic limit.
	
		\bibliography{BiblioEPT_Semi}
	
\end{document}